\documentclass[a4paper,11pt]{article}
\usepackage{hyperref}

\def\be{\begin{equation}}
\def\te{\end{equation}}
\def\ee{\end{equation}}
\def\ba{\begin{eqnarray}}
\def\bea{\begin{eqnarray}}
\def\nn{\nonumber\\}
\def\tea{\end{eqnarray}}
\def\ea{\end{eqnarray}}
\def\eea{\end{eqnarray}}
\begin{document}

\title{Quantum Field Approach to Relativistic Turbulence }

\author{Esteban Calzetta\\
Universidad de Buenos Aires, Facultad de Ciencias Exactas y Naturales,\\ Departamento de Física, Argentina,\\ y CONICET - Universidad de Buenos Aires,\\ Instituto de Física de Buenos Aires (IFIBA), Argentina\thanks{calzetta@df.uba.ar}}

\maketitle

\begin{abstract}{The goal of this work is to apply quantum field theory methods to a typical relativistic hydrodynamic model. We shall take as representative model an Israel-Stewart framework, where the energy-momentum conservation laws for a conformal invariant fluid are supplemented by a Cattaneo-Maxwell equation for its viscous energy-momentum tensor. In these models the viscous energy-momentum tensor relaxes to its Landau-Lifshitz value on a finite time scale. We assume the parameters of the model depend on the speed of light $c$ in such a way that as $c\to\infty$ the fluid becomes an incompressible fluid obeying the Navier-Stokes equations. We show that because the relativistic model must be hyperbolic, each mode behaves as an overdamped oscillator with two decay constants, one of which diverges as $c\to\infty$. Thus for finite $c$ there are two basic patterns, one which contains only slowly decaying  modes, and reproduces Kolmogorov turbulence when $c\to\infty$, and another made of fast decaying modes only. Each flow pattern is associated to definite relations among the velocity and viscous energy tensor correlators.}\end{abstract}

%\keyword{relativistic hydrodynamics; quantum field theory; turbulence} 

%\begin{document}

\section{Introduction}

The goal of this work is to apply quantum field theory methods to a typical relativistic hydrodynamic model \cite{RZ13,RR19}. We focus on which differences between relativistic and nonrelativistic turbulence follow from the fact that the former must be a hyperbolic theory, while the latter is parabolic.

In spite of its relevance for relativistic heavy ion collisions \cite{K08,FW11,CR11,BBSV14,F13,AKLN14}, cosmology \cite{GC02,MT03,MT04,Crespo,Crespo2,nikschlesigl18}
 and the astrophysics of compact objects \cite{BN11,LBM13,Andrade,Saha}, and significant progress in the numerical simulation of special and general relativistic turbulent flows \cite{Zrake,Rezzolla1,Rezzolla2,Celora2,Celora3}, it is fair to say that the theory of relativistic turbulence is not anywhere close to the depth of its non relativistic counterpart, epitomized by the Kolmogorov 1941 (K41) theory \cite{Chandra54,MY71,MY75,Frisch}.

In the same way that the study of non-relativistic turbulence has its point of departure in the Navier-Stokes equations (NSE), we shall study a definite set of equations describing a real relativistic  fluid. For an ideal relativistic fluid these equations are the relativistic Euler equations, namely the conservation equations for the energy-momentum tensor (EMT). For a real fluid, no model has obtained the consensus attached to the NSE in the non-relativistic domain. 

For concreteness, we shall adopt a Israel - Stewart type model \cite{israel76,IsSte76,IsSte79a,IsSte79b,IsSte80,israel88}. In these models, the conservation equations for the EMT are supplemented by a Cattaneo - Maxwell \cite{Max67,Catt48,Catt58,JosPrez89}  equation for the viscous EMT, which is regarded as an independent degree of freedom on its own right. The model is such that for long times it relaxes to Landau-Lifshitz relativistic hydrodynamics \cite{LL6}, while being causal and stable \cite{HL83,SM22,BRD23,HK24,WP24,GDN24,BhaMiRo24}. For simplicity, we assume that the energy density diverges as $c^2$ when $c\to\infty$; the model then becomes an incompressible Newtonian fluid obeying the NSE in the non relativistic limit \cite{FOz08}. We shall further discuss this assumption below. This model is arguably too simple to be useful in actual applications, such as the generation of gravitational waves from fluid fluctuations in the early Universe \cite{Juampi}, but it will allow us to discuss the main conceptual issues without getting too deep into merely technical difficulties. 

To stir the fluid, we shall assume that both the EMT conservation equations and the Cattaneo - Maxwell equations are subject to stochastic Gaussian noises \cite{Novikov1,Novikov,Poursina}. 

To analyze the model, we shall consider it as a deformation of the NSE and study it perturbatively, taking $1/c^2$ as the small parameter. By construction, at leading order the model reduces to the NSE. A similar strategy was recently used to study turbulence in Non-Newtonian fluids\cite{EC25}.

The non relativistic limit of a causal fluid is singular \cite{vanDyke}, because the hyperbolic equations of motion of the causal fluid become the parabolic NSE.

To see the implications of this fact, let us recall the limit of a $LRC$ electric circuit when $L\to 0$. The equations for this model, for $q$ the capacitor charge and $j$ the current, are

\bea
\dot q-j&=&0\nn
L\dot j+Rj+\frac qC&=& v
\label{LRC}
\tea
where $v$ is a random delta-correlated Gaussian noise with zero mean and amplitude determined by the fluctuation - dissipation theorem

\be
\left\langle v\left(t\right)v\left(t'\right)\right\rangle=2Rk_BT\delta\left(t-t'\right)
\label{FDT}
\te
As in the fluid model, the first equation is a conservation law, namely that of charge, and the second equation describes the relaxation of the current to its steady value. Note that there is no noise term in the first equation (\ref{LRC}), because charge conservation is assumed to be exact and the model is linear \cite{LandauSP}. For the fluid model, even if no stochastic driving is included in the EMT conservation equation, stochasticity is generated through nonlinearities. 

For $L$ finite but small ($L\le CR^2/4$) the model represents an overdamped oscillator with two definite damping rates. As $L\to 0$, one of them converges to the damping rate $1/RC$ of a $RC$ circuit, while the other diverges. This vanishing mode affects the dynamics as long as $L$ remains finite. Due to the interference between the two modes, the response to an impulse of the $LRC$ circuit is continuous, while that of the $RC$ circuit is not. 

All of this is trivial, but the point is that after all the complications are cleared away, relativistic hydrodynamics displays essentially the same behavior. When  $c^2$ is large but finite, each mode of the fluid behaves as an overdamped oscillator. When $c^2\to\infty$, one of the decaying rates converges to its NSE value, while the other diverges. As we shall show in the following, these doubling of the decay rates leads to the existence of two basic flow patterns, characterized by the presence of only slowly decaying or fast decaying modes, which in turn is possible if the velocity and viscous energy-momentum tensor obey definite relations. %One is a ``dressed'' Kolmogorov cascade, with the same exponents and a scale invariant energy flux as in K41 theory, while the other is an independent, fast decaying flow which, under certain scaling relations to be discussed below (see eq. (\ref{scalingsmas})), may organize itself into an entropy cascade \cite{Eyink18b,EC1}.  To the best of our knowledge, the detailed description of the second cascade is a new result.

To delve into these issues we shall translate the problem into a field theory one by using the Martin-Siggia-Rose (MSR) formalism \cite{msr-73,dedo-76,DeDomMar79,DeDGia06,FoNeSte77,GBF81,Ph77}. There is a vast literature on the application of MSR to Kolmogorov turbulence \cite{LB87,E94,eyink96,T97,Kromnes97,Kromnes02}. Moreover, earlier work by Kraichnan, Wyld and Lee is known to be equivalent to it \cite{Kr61,Wyld61,Lee65,BSMcC}. For its application to causal relativistic models see \cite{Nahuel20,MGKC}. 

In spite of this body of work, one also finds in the literature \cite{Orszag70,K71,G98,Tsin09,MC23,MC24}  reservations about whether field theory, and specifically the MSR approach, can ultimately crack ``the problem of turbulence'' \cite{Eckert,Nelkin,Sreenivasan}. The importance of this criticism should not be underestimated, because if such is the case, not only the MSR approach, but the whole of non equilibrium field theory \cite{Ram07,calhu08,kam-11} would become suspect. Thus, another motivation for this work is to provide elements for this debate, freely acknowledging that we do not foresee a definitive conclusion for it anytime soon.

Since one of the flow patterns contain only the modes which converge to NSE solutions as $c\to\infty$, it is natural to ask whether in this limit this pattern corresponds to the usual Kolmogorov cascade. To obtain this cascade, whose existence in relativistic turbulence has been shown in \cite{FOz10}, we must assume further scaling relations, which amount to the Kolmogorov choice of the energy dissipation rate as the single determinant dimensionful parameter in the fully developed turbulence regime. This leads to scale invariant solutions are associated to a direct energy cascade \cite{Kraichnan67,Wal92,ED18b,E18,Biferale,Falkovich}. 

The remainder of the paper is organized as follows. In next section (\ref{MSR}) we present an overview of the MSR approach; this overlaps to some extent with references \cite{Nahuel20,MGKC,EC1}, and it is included only for completeness and to fix the notation. Ref. \cite{Nahuel20} introduces the MSR approach and functional techniques to study fluctuating relativistic fields, but the emphasis is on thermal, rather than turbulent, fluctuations. Ref. \cite{MGKC} considers a relativistic hydrodynamic model derived from kinetic theory, which contains a propagating tensor mode, and only discusses fluctuations  of this mode, when driven by thermal noise. Ref. \cite{EC1} pursues the discussion of fluctuations of this tensor mode, not using the functional approach, but rather a relativistic K\'arm\'an-Howarth equation \cite{vKH}.

We present the model to be used in this paper in section (\ref{CM}), first in full relativistic form. The model is built by regarding the viscous part of the EMT as a non hydrodynamic degree of freedom on its own, and then providing an equation of motion which describes its relaxation to its Landau-Lifshitz value. We then develop each field and the equations of motion in inverse powers of $c$. As we shall see, it is possible to discard subdominant terms in the non relativistic limit, thus obtaining a simpler model which is nevertheless causal and converges to the NSE for an incompressible fluid as $c\to\infty$. In particular, the model we shall discuss is much simpler than those presented in refs \cite{Nahuel20,MGKC}. We stress that once the final equations of the model are found, eqs. (\ref{eq11}) and (\ref{eq21}) below, we shall discuss them on their own right, setting aside their derivation from the large $c$ expansion of the full relativistic theory. We believe that presenting this model, which is relatively simple but retains the basic characteristics of causality, Lorentz invariance and stability, is one of the main results of this work.

In the following section (\ref{largec}) we show how to write down the causal propagators and  the correlation functions of the model in terms of the self energies and noise kernels appearing in the 2PIEA. Under the Markovian approximation eq. (\ref{Markovianapproximation}), we shall find that, as in the $LRC$ circuit discussed above, for each mode there are two different decay rates, one that remains finite and another that diverges when $c^2\to\infty$. For further discussion of this approximation we refer the reader to \cite{Canet22}. We then discuss under which conditions the relativistic model makes contact with the non-relativistic K41 theory, and the analogous conditions which lead to a flow pattern containing fast decaying modes only.

We conclude the paper with some brief final remarks, pointing out some avenues for further development. 

In the analysis of self similar solutions for the slowly decaying flow, we find that supplementary conditions are necessary to point out a preferred solution; in this case, the new condition is the scale invariance of the energy flow. As a motivation for further work, we discuss in Appendix A whether the hypothesis of a scale invariant entropy flow \cite{Eyink18b} could help singling out a self similar solution for the fast decaying flow.

\section{Functional methods and the Martin - Siggia and Rose approach}
\label{MSR}

In this section, we present an overview of non equilibrium field theory methods and the MSR approach. It is included for completeness only, see \cite{Nahuel20,MGKC} for further details.

\subsection{The two particle irreducible effective action}

Let us start with a summary of the functional MSR method. Let us first consider the functional aspect. We are working with a theory of fields $X^{\alpha}$, where $\alpha$ accounts for both discrete and continuous indexes. We assume a cubic action

\bea
S&=&S_q+S_s\nn
S_q&=&\frac12 D_{\alpha\beta}X^{\alpha}X^{\beta}\nn
S_c&=&\frac16 D_{\alpha\beta\gamma}X^{\alpha}X^{\beta}X^{\gamma}
\tea
We apply Einstein's convention to both continuous and discrete indexes and assume that the coefficients in the action are totally symmetric. The generating functional

\be
e^{iW}=\int DX\;e^{i\left(S+S_s\right)}
\te
where $S_s$ is the source action

\be
S_s=P_{\alpha}X^{\alpha}+\frac12P_{\alpha\beta}X^{\alpha}X^{\beta}
\te 
The derivatives of the generating functional define the mean fields and the propagators

\bea
\frac{\delta W}{\delta P_{\alpha}}&=&\left\langle X^{\alpha}\right\rangle\equiv \bar X^{\alpha}\nn
\frac{\delta W}{\delta P_{\alpha\beta}}&=&\frac12\left\langle X^{\alpha}X^{\beta}\right\rangle\equiv\frac12\left[ G^{\alpha\beta}+\bar X^{\alpha}\bar X^{\beta}\right]
\tea 
The two particle irreducible effective action (2PIEA) is the full Legendre transform

\be
\Gamma =W -P_{\alpha}\bar X^{\alpha}-\frac12P_{\alpha\beta}\left[ G^{\alpha\beta}+\bar X^{\alpha}\bar X^{\beta}\right]
\te 
whereby we get the equations of motion

\bea
\frac{\delta \Gamma}{\delta \bar X^{\alpha}}&=&-P_{\alpha}-P_{\alpha\beta}\bar X^{\beta}\nn
\frac{\delta \Gamma}{\delta G^{\alpha\beta}}&=&-\frac12P_{\alpha\beta}
\tea
From now on we shall set $P_{\alpha\beta}=0$. The 2PIEA has the structure

\be
\Gamma=S\left[\bar X\right]+\frac12\left[D_{\alpha\beta}+D_{\alpha\beta\gamma}\bar X^{\gamma}\right]G^{\alpha\beta}-\frac i2\;{\rm{Tr}}\;\log\left[ G\right]+\Gamma_Q
\te
$\Gamma_Q$ is the sum of all two particle irreducible Feynman graphs with the propagators $G^{\alpha\beta}$ in internal lines and vertices extracted from $S_c$. We shall adopt the lowest order approximation

\be
\Gamma_Q=\frac i{2}\left\langle S_c^2\right\rangle_{2PI}
\label{settingsun}
\te
$\Gamma_Q$  does not depend on the mean fields, so the equations of motion take the form

\be
D_{\alpha\beta}\bar X^{\beta}+\frac12 D_{\alpha\beta\gamma}\left[\bar X^{\beta}\bar X^{\gamma}+G^{\beta\gamma}\right]=-P_{\alpha}
\label{MF}
\te

\be
\frac12\left[D_{\alpha\beta}+D_{\alpha\beta\gamma}\bar X^{\gamma}\right]-\frac i2\left[G^{-1}\right]_{\alpha\beta}+\frac{\delta \Gamma_Q}{\delta G^{\alpha\beta}}=0
\te
The second equation may be rewritten as

\be
\left[D_{\alpha\beta}+D_{\alpha\beta\gamma}\bar X^{\gamma}+2\frac{\delta \Gamma_Q}{\delta G^{\alpha\beta}}\right]G^{\beta\delta}=i\mathbf{1}^{\delta}_{\alpha}
\label{props1}
\te 
where $\mathbf{1}^{\delta}_{\alpha}$ is the identity operator in the space of the fields $X^{\alpha}$. 

\subsection{Higher correlations}

We may investigate higher correlations by means of the identity, valid for any operator $\mathcal {O}$

\be 
\frac{\delta }{\delta P_{\alpha}}\left\langle \mathcal {O}\right\rangle=i\left[\left\langle \mathcal {O}X^{\alpha}\right\rangle-\left\langle \mathcal {O}\right\rangle\bar X^{\alpha}\right]
\te
We envisage throughout a situation where the mean fields vanish on-shell, namely when the $P_{\alpha}=0$, whereby 

\be 
\frac{\delta G^{\beta\gamma}}{\delta P_{\alpha}}=i\left\langle X^{\alpha}X^{\beta}X^{\gamma}\right\rangle\equiv iH^{\alpha\beta\gamma}
\te
The variation of equation (\ref{MF}) wrt $P_{\delta}$ shields a set of von K\'arm\'an-Howarth like identities \cite{vKH}, which on shell read

\be
D_{\alpha\beta}G^{\beta\delta}+\frac12 D_{\alpha\beta\gamma}H^{\beta\gamma\delta}=i\mathbf{1}^{\delta}_{\alpha}
\label{vKH}
\te
The variation of eq. (\ref{props1}) yields

\be
\left[D_{\alpha\beta}+D_{\alpha\beta\gamma}\bar X^{\gamma}+2\frac{\delta \Gamma_Q}{\delta G^{\alpha\beta}}\right]H^{\beta\delta\epsilon}=-\left[D_{\alpha\beta\gamma}G^{\gamma\epsilon}+2\frac{\delta^2 \Gamma_Q}{\delta G^{\alpha\beta}\delta G^{\xi\eta}}H^{\xi\eta\epsilon}\right]G^{\beta\delta}
\label{higherprops}
\te 
Using eq. (\ref{props1}) again, this becomes

\be 
H^{\zeta\delta\epsilon}=\left(i\right)G^{\zeta\alpha}\left[D_{\alpha\beta\gamma}G^{\gamma\epsilon}+2\frac{\delta^2 \Gamma_Q}{\delta G^{\alpha\beta}\delta G^{\xi\eta}}H^{\xi\eta\epsilon}\right]G^{\beta\delta}
\label{higherprops2}
\te 
We may regard eq. (\ref{higherprops2}) as a set of Bethe-Salpeter equations for the three point functions \cite{Kromnes78}, subject to the consistency conditions eqs. (\ref{vKH}).

If we neglect the second variation of $\Gamma_Q$ in the Bethe-Salpeter equation, then

\be
H^{\zeta\delta\epsilon}=iG^{\zeta\alpha}D_{\alpha\beta\gamma}G^{\beta\delta}G^{\gamma\epsilon}
\label{higherprops3}
\te 
which is in fact Kraichnan's Direct Interaction Approximation (DIA) 
\cite{Kr59}
and

\be
D_{\alpha\beta}G^{\beta\delta}+\frac i2 D_{\alpha\beta\gamma}G^{\beta\zeta}D_{\zeta\xi\eta}G^{\xi\gamma}G^{\eta\delta}=i\mathbf{1}^{\delta}_{\alpha}
\label{vKH2}
\te
Using eq. (\ref{props1}) this becomes

\be
\left[\frac i2 D_{\alpha\eta\gamma}G^{\eta\zeta}G^{\gamma\xi}D_{\zeta\xi\beta}-2\frac{\delta \Gamma_Q}{\delta G^{\alpha\beta}}\right]G^{\beta\delta}=0
\te 
Since $G^{\beta\delta}$ is invertible, the expression within brackets must be identically zero. This is actually what one gets as the variation of the effective action under the approximation eq. (\ref{settingsun}), showing that neglecting the Hessian of $\Gamma_Q$ in eq. (\ref{higherprops2}) is consistent at this level of accuracy.

\subsection{Martin, Siggia and Rose (MSR)}

The MSR framework addresses a situation where the physical fields $V^j$ obey equations

\be
D^j_kV^k+\frac12D^j_{kl}V^kV^l=f^j
\te 
The $f^j$ are Gaussian random forces with zero mean and self-correlation $\left\langle f^jf^k\right\rangle=N_0^{jk}$. They are necessary to stir the fluid. They must be random, since the expectation value of the physical fields would not be zero  otherwise.

To derive these from a variational principle one includes a corresponding number of Lagrange multipliers $A_j$ and writes an action

\be
S=A_j\left[D^j_kV^k+\frac12D^j_{kl}V^kV^l\right]+\frac i2A_jN_0^{jk}A_k
\te
The imaginary term arises from the average over the random forces. So

\bea
&&S_q=A_jD^j_kV^k+\frac i2A_jN_0^{jk}A_k\nn
&&S_c=\frac12A_jD^j_{kl}V^kV^l
\tea
The auxiliary and physical fields together make up the $X^{\alpha}$ fields we discussed in the previous section. 

To summarize, we have obtained an action where the degrees of freedom are the original physical fields and an equal number of auxiliary fields. The problem is classical, but the presence of noise justifies borrowing tools from quantum field theory \cite{calhu08} to analyze it. To see this, let $\nu_0$ be a quantity specifying the overall scale of the noise, $N_0^{jk}=\nu_0\tilde N_0^{jk}$. If we rescale the auxiliary fields $A_j\to \nu_0^{-1}\tilde A_j$, we see that the action acquires a global factor of $\nu_0^{-1}$. Thus, the noise scale $\nu_0$ plays the role of ``$\hbar$'', and conventional field theory techniques such as the loop expansion become an expansion in powers of the noise.

On-shell the expectation value of the physical fields (the ``mean fields'') vanish, as well as the expectation values of a product of any number of auxiliary fields alone. 

Recall that
\be
\left(\begin{array}{cc}\left[G^{-1}\right]_{VVjk}&\left[G^{-1}\right]^k_{VAj}\\\left[G^{-1}\right]^j_{AVk}&\left[G^{-1}\right]^{jk}_{AA}\end{array}\right)
\left(\begin{array}{cc}G^{kl}_{VV}&G^k_{VAl}\\G^l_{AVk}&G_{AAkl}\end{array}\right)=\delta\left(t-t'\right)\left(\begin{array}{cc}\Delta^{l}_j&0\\0&\Delta^j_l\end{array}\right)
\te
On-shell $G_{AA}=0$, so 

\bea
\left[G^{-1}\right]_{VVjk}G^{kl}_{VV}+\left[G^{-1}\right]^k_{VAj}G^k_{AVl}&=&\Delta^{l}_j\nn
\left[G^{-1}\right]_{VVjk}G^k_{VAl}&=&0\nn
\left[G^{-1}\right]^j_{AVk}G^{kl}_{VV}+\left[G^{-1}\right]^{jk}_{AA}G^l_{AVk}&=&0\nn
\left[G^{-1}\right]^j_{AVk}G^k_{VAl}=\Delta^{j}_l
\tea 
Therefore, on-shell

\bea 
\left[G^{-1}\right]_{AV}&=&\left[G_{VA}\right]^{-1}\nn
\left[G^{-1}\right]_{VA}&=&\left[G_{AV}\right]^{-1}\nn
\left[G^{-1}\right]_{VV}&=&0
\tea
and

\be
G^{jk}_{VV}=-G^j_{VAl}\left[G^{-1}\right]^{lm}_{AA}G^k_{AVm}
\label{Hadamard}
\te
We can then write the equations of motion on-shell. The mean field equations become

\bea
&&D^j_{kl}G^{kl}_{VV}=0\nn
&&D^j_{kl}G^{l}_{AVj}=0
\label{MSRMF}
\tea
and the Schwinger-Dyson equations 

\bea
&&\frac i2N_0^{jk}-\frac i2\left[G^{-1}\right]_{AA}^{jk}+\frac{\delta \Gamma_Q}{\delta G_{AAjk}}=0\nn
&&\frac12D^j_k-\frac i2\left[G_{VA}\right]^{-1j}_k+\frac{\delta \Gamma_Q}{\delta G^k_{AVj}}=0\nn
&&\frac12D^k_j-\frac i2\left[G_{AV}\right]^{-1k}_j+\frac{\delta \Gamma_Q}{\delta G^j_{VAk}}=0\nn
&&\frac{\delta \Gamma_Q}{\delta G_{VV}^{jk}}=0
\tea
The first of these implies that

\be
\left[G^{-1}\right]_{AA}^{jk}=N_0^{jk}+N^{jk}
\te
where

\be
N^{jk}=-2i\frac{\delta \Gamma_Q}{\delta G_{AAjk}}
\label{noise}
\te
are the noise kernels. We are interested in a regime where $N$ dominates over $N_0$.

We also write down the three point correlations of physical fields  to lowest order in $S_c$, namely the closure relation \cite{FGB94,FOK,FKO24,Zhou}

\be
\left\langle V^jV^kV^l\right\rangle=i\left\langle V^jV^kV^l\;S_c\right\rangle_{2PI}
\label{VKHMSR}
\te

\section{Conformal fluids and their non-relativistic limit}\label{CM}

In this section we shall present a fully relativistic model, and then proceed to a systematic expansion in inverse powers of the speed of light $c$. We shall keep only the most relevant terms at leading and next to leading orders.

We consider a relativistic fluid of massless particles. At the macroscopic level, the theory is described by the energy-momentum tensor (EMT) $T^{\mu\nu}$. Since the fluid is massless, the chemical potential vanishes. Adopting the Landau prescription for the four velocity $u^{\mu}$ and the energy density $\rho$  \cite{LL6}

\be
T^{\mu}_{\nu}u^{\nu}=-\rho u^{\mu}
\te
Since $T^{\mu\nu}$ is traceless, we are led to write

\be
T^{\mu\nu}=\rho\left[u^{\mu}u^{\nu}+\frac13\bar{\Delta}^{\mu\nu}+\Pi^{\mu\nu}\right]
\label{tmunu}
\te
where

\be
\bar{\Delta}^{\mu\nu}=\eta^{\mu\nu}+u^{\mu}u^{\nu}
\te 
and

\be
\Pi^{\mu}_{\nu}u^{\nu}=\Pi^{\mu}_{\mu}=0
\label{TT}
\te
The energy density $\rho$ is related to the temperature $T$ through

\be
\rho=\sigma_{SB} T^4
\te
for some constant $\sigma_{SB}$. Observe that both $u^{\mu}$ and $\Pi^{\mu\nu}$ are dimensionless. Then the EMT conservation laws are 

\be
Q=\frac1{\rho}\rho_{,\nu}u^{\nu}+\frac43u^{\nu}_{,\nu}+\Pi^{\mu\nu}u_{\mu,\nu}=0
\label{cons0}
\te
and

\be
Q^{\mu}=\frac1{\rho}\rho_{,\nu}\left[\bar{\Delta}^{\mu\nu}+3\Pi^{\mu\nu}\right]+4u^{\mu}_{,\nu}u^{\nu}+3\bar{\Delta}^{\mu}_{\rho}\Pi^{\rho\nu}_{,\nu}=0
\label{consj}
\te 
These equations are complemented by  the Cattaneo-Maxwell equation \cite{Max67,Catt48,Catt58,JosPrez89} 

\bea
Q^{\rho\sigma}&=&\lambda u^{\nu}\bar{\Delta}^{\rho}_{\rho'}\bar{\Delta}^{\sigma}_{\sigma'}\Pi^{\rho'\sigma'}_{,\nu}+\frac1{\ell}\Pi^{\rho\sigma}+\frac18\sigma^{\rho\sigma}\nn
&=&\lambda \left[u^{\nu}\Pi^{\rho\sigma}_{,\nu}-\left(u^{\rho}\Pi^{\lambda\sigma}+u^{\sigma}\Pi^{\rho\lambda}\right)u^{\nu}u_{\lambda,\nu}\right]+\frac1{\ell}\Pi^{\rho\sigma}+\frac18\sigma^{\rho\sigma}=0
\label{CM2}
\tea
$\lambda$ and $\ell$ are constants, which we take as free parameters. $\lambda$ is dimensionless, while $\ell$ has dimensions of length. Here

\be
\sigma^{\rho\sigma}=\left[\bar{\Delta}^{\rho\mu}\bar{\Delta}^{\sigma\nu}+\bar{\Delta}^{\rho\nu}\bar{\Delta}^{\sigma\mu}-\frac23\bar{\Delta}^{\rho\sigma}\bar{\Delta}^{\mu\nu}\right]u_{\mu,\nu}
\te
is the covariant form of the shear tensor. Eq. (\ref{CM2}) ensures positive entropy production, given the entropy flux (compare with \cite{Obu49})

\be
S^{\mu}=su^{\mu}
\te
with entropy density 

\be
s=\frac43\sigma_{SB}T^3e^{-\frac32\lambda\Pi^{\mu\nu}\Pi_{\mu\nu}}
\label{sdens}
\te 
The entropy production is nonnegative

\be
0\le S^{\mu}_{,\mu}=\frac 3{\ell}s\Pi^{\rho\sigma}\Pi_{\rho\sigma}
\te
This ensures the stability of the model \cite{HL83,SM22,BRD23,HK24,WP24,GDN24,BhaMiRo24}.

\subsection{Non-relativistic expansion of the fields}

Eqs. (\ref{cons0}), (\ref{consj}) and (\ref{CM2}) define the model in 
fully relativistic form. We now proceed to expand the relevant fields and equations in inverse powers of the speed of light $c$ \cite{FOz08}. We first consider the fields.

\subsubsection{Expansion of the velocity and its derivatives}

We shall now consider the non relativistic limit. We write explicitly $x^0=ct$ and
\be
u^{\mu}=u^0\left( 1,v^k/c\right);\;\;u^0=\frac1{\sqrt{1-v^2/c^2}} 
\te 
We shall call

\be
DX=X_{,t}+v^kX_{,k}
\te
Observe that

\be
\bar{\Delta}^{\mu}_{\nu}=\left( \begin{array}{cc}-v^2/c^2&v^k/c\left(1+\frac{v^2}{c^2}\right)\\-v_j/c\left(1+\frac{v^2}{c^2}\right)&\delta^{jk}+\frac1{c^2}v^jv^k\end{array}\right)+O\left(\frac1{c^3}\right)
\te
Now

\bea
&&u^{\mu}_{;\mu}=\frac{u^0}cv^k_{,k}+\frac1cDu^0\nn
&=&\frac1c\left(1+\frac{v^2}{2c^2}\right)v^k_{,k}+\frac{1}{2c^3}D\left(v^2\right)
\tea

\bea
&&u_{\mu;\nu}=\left(\begin{array}{cc}\frac{-1}c\left(u^0\right)_{,t}&-u^0_{,k}\\\frac{1}{c^2}\left(u^0v^j\right)_{,t}&\frac{1}c\left(u^0v^j\right)_{,k}\end{array}\right)\nn
&\approx&\left(\begin{array}{cc}\frac{-1}{2c^3}\left(v^2\right)_{,t}&\frac{-1}{2c^2}v^2_{,k}-\frac3{4c^4}v^2\left(v^2\right)_{,k}\\\frac{1}{c^2}v^j_{,t}+\frac1{2c^4}\left(v^2v^j\right)_{,t}&\frac1cv^j_{,k}+\frac1{2c^3}\left(v^2v^j\right)_{,k}\end{array}\right)+O\left(\frac1{c^5}\right)
\tea
which is consistent with $u^{\mu}u_{\mu;\nu}=0$
\be
u_{\mu;\nu}u^{\nu}=\left(\begin{array}{c}\frac{-1}{2c^3}Dv^2\\\frac1{c^2}\left(1+\frac{v^2}{c^2}\right)Dv_j+\frac{v_j}{2c^4}D\left(v^2\right)\end{array}\right)+O\left(\frac1{c^5}\right)
\te

\bea
&&u_{\mu,\rho}\bar{\Delta}^{\rho}_{\nu}=\left(\begin{array}{cc}\frac{-1}{2c^3}\left(v^2\right)_{,t}&\frac{-1}{2c^2}v^2_{,l}-\frac3{4c^4}v^2\left(v^2\right)_{,l}\\\frac{1}{c^2}v^j_{,t}+\frac1{2c^4}\left(v^2v^j\right)_{,t}&\frac1cv^j_{,l}+\frac1{2c^3}\left(v^2v^j\right)_{,l}\end{array}\right)\nn
&&\left( \begin{array}{cc}-v^2/c^2&v^k/c\left(1+\frac{v^2}{c^2}\right)\\-v_l/c\left(1+\frac{v^2}{c^2}\right)&\delta^{lk}+\frac1{c^2}v^lv^k\end{array}\right)\nn
&=&\left(\begin{array}{cc}\frac1{2c^3}v^lv^2_{,l}&-\frac1{2c^2}v^2_{,k}-\frac{1}{2c^4}v^kDv^2-\frac3{4c^4}v^2\left(v^2\right)_{,k}\\-\frac1{c^2}v^lv^j_{,l}-\frac1{c^4}v^2Dv^j-\frac1{2c^4}v^l\left(v^2v^j\right)_{,l}&\frac1cv^j_{,k}+\frac1{2c^3}\left(v^2v^j\right)_{,k}+\frac1{c^3}v^kDv^j\end{array}\right)
\tea
We only write down

\bea
&&\sigma_{jk}=\frac1c\left[v_{j,k}+v_{k,j}-\frac23v^l_{,l}\delta_{jk}\right]+\frac1{2c^3}\left[\left(v^2v_j\right)_{,k}+\left(v^2v_k\right)_{,j}-\frac23\left(v^2v^l\right)_{,l}\delta_{jk}\right]\nn
&+&\frac1{c^3}\left[v_kDv_j+v_jDv_k-\frac13Dv^2\delta_{jk}\right]-\frac2{3c^3}\left[v_jv_kv^l_{,l}-\frac12v^lv^2_{,l}\delta_{jk}\right]
\tea
Observe that $\sigma^j_j=-\sigma^0_0\not=0$.

\subsubsection{Expansion of the viscous energy-momentum tensor}
The most general form of the viscous EMT compatible with the constraints eq. (\ref{TT}) is

\bea
\Pi^{\mu\nu}&=&\left( \begin{array}{cc}\Pi_{lm}v^lv^m/c^2&\Pi_{kl}v^l/c\\\Pi_{jm}v^m/c&\Pi_{jk}\end{array}\right) +\frac{\Pi_{lm}v^lv^m/c^2}{3-v^2/c^2}\left( \begin{array}{cc}v^2/c^2&v^k/c\\v^j/c&\delta_{jk}\end{array}\right)\nn
&\approx&\left( \begin{array}{cc}\Pi_{lm}v^lv^m/c^2&\Pi_{kl}v^l/c\\\Pi_{jm}v^m/c&\Pi_{jk}+\frac1{3c^2}\Pi_{lm}v^lv^m\delta_{jk}\end{array}\right) 
\tea
where $\Pi^j_j=0$. Therefore,

\be
\Pi^{\mu\nu}_{,\nu}\approx\left( \begin{array}{c}\frac1c\left[\left(\Pi_{kl}v^l\right)_{,k}+\frac1{c^2}\left(\Pi_{lm}v^lv^m\right)_{,t}\right]\\\Pi_{jk,k}+\frac1{c^2}\left[\left(\Pi_{jm}v^m\right)_{,t}+\frac1{3}\left(\Pi_{lm}v^lv^m\right)_{,j}\right]\end{array}\right) 
\te

\subsubsection{The energy density}

In the non relativistic limit, we may extract a mass density from the energy density by writing

\be
\rho=\mu c^2+\epsilon
\te
We make the assumption that $\mu$ remains finite as $c\to\infty$ \cite{FOz08}. Then

\bea
&&\rho_{;\mu}u^{\mu}=\frac{u^0}cD\rho\nn
&\approx&\frac{1}c\left(1+\frac{v^2}{2c^2}\right)D\rho,
\tea
and

\be
\frac1{\rho}\rho_{,\nu}=\frac1{\mu\;c^2}\epsilon_{,\nu}\left[1-\frac{\epsilon}{\mu\; c^2}\right]\approx\frac1{\mu\; c^2}\epsilon_{,\nu}
\te
Assuming that $\mu$ remains finite is a tremendous simplification, because then the fluid becomes incompressible as $c\to\infty$. It would be more satisfactory, however, not to depend on this hypothesis and deal rather with a compressible model. Nevertheless, even in a compressible model the velocity field may be decomposed into compressible and incompressible components, and one can identify a ``sonic scale'' where the turbulent velocity matches the thermal velocity (see \cite{McKO07}). If the Reynolds number at this scale is large enough, the incompressible component will dominate at smaller scales, and the fluid will be approximately incompressible there.

\subsection{Expansion of the equations of motion}

We now collect these results in a set of independent equations of motion..

Since equations (\ref{consj}) are not independent because $u_{\mu}Q^{\mu}=0$ identically, it is enough to consider the spatial terms $Q^j$. For the same reason, it is enough to consider the $Q^{jk}$ equations. 

We see that $u^j_{,\nu}u^{\nu}$ is a term of order at least $1/c^2$. In order not to be subdominant with respect to the other terms in the momentum conservation equation, we require $\rho_{,\nu}/\rho$ and $\Pi^{\mu\rho}_{,\rho}$ to scale accordingly. This requires the mass density $\mu=$ constant 
\cite{FOz08} and

\be
\Pi^{ij}=\frac1{c^2}p^{ij},
\te
where $p^{ij}$ has dimensions of velocity squared. In the following 
we proceed with the analysis of the equations of motion. 

\subsubsection{Energy conservation}

(Eq. (\ref{cons0})) becomes

\be
0=\frac1{\mu c^3}D\epsilon+\frac{4}{3}\left[\frac1c\left(1+\frac{v^2}{2c^2}\right)v^k_{,k}+\frac{1}{2c^3}D\left(v^2\right)\right]+\frac1{c^3}p^{jk}u_{j,k}
\te

\subsubsection{Momentum conservation}

From eq. (\ref{consj})

\bea
&&0=4\frac1{c^2}\left(1+\frac{v^2}{c^2}\right)Dv_j+\frac{v_j}{2c^4}D\left(v^2\right)\nn
&+&\frac1{\mu\;c^2}\epsilon_{,j}\left[1-\frac{\epsilon}{\mu\; c^2}\right]+\frac1{\mu\; c^4}v^jD\epsilon+\frac3{\mu\;c^4}p^{jk}\epsilon_{,k}\nn
&+&\frac3{c^2}p_{jl,l}+\frac3{c^4}\left[\left(p_{jm}v^m\right)_{,t}+\frac1{3}\left(p_{lm}v^lv^m\right)_{,j}\right]+\frac3{c^4}v^jv^kp_{kl,l}-\frac3{c^4}v^j\left(p_{kl}v^l\right)_{,k}
\tea

\subsubsection{The Cattaneo-Maxwell equation }

From eq. (\ref{CM2}),
keeping terms up to $1/c^3$

\bea
&&0=\frac1{8c}\left[v_{j,k}+v_{k,j}-\frac23v^l_{,l}\delta_{jk}\right]+\frac1{16c^3}\left[\left(v^2v_j\right)_{,k}+\left(v^2v_k\right)_{,j}-\frac23\left(v^2v^l\right)_{,l}\delta_{jk}\right]\nn
&+&\frac1{8c^3}\left[v_kDv_j+v_jDv_k-\frac13Dv^2\delta_{jk}\right]-\frac1{12c^3}\left[v_jv_kv^l_{,l}-\frac12v^lv^2_{,l}\delta_{jk}\right]\nn
&+&\frac1{\ell\; c^2}\left[p_{jk}+\frac1{3c^2}p_{lm}v^lv^m\delta_{jk}\right]+\frac{\lambda}{c^3}Dp_{jk}
\tea
Observe that this equation is consistent with $p^k_k=0$ to the given order. On dimensional grounds, we expect $\ell\approx \eta/c$, where $\eta$ is the kinetic viscosity. We write

\be
\ell=\frac{32}3\frac{\eta}c
\te
to match the NSE. 

In Kolmogorov turbulence the velocity at scale $\delta$ scales as $\left(\varepsilon\delta\right)^{1/3}$, reaching $\left(\eta\varepsilon \right)^{1/4}$ at the Kolmogorov scale $\delta_K\approx \left(\eta^3/\varepsilon\right)^{1/4}$ \cite{FrischOrszag}. The several relativistic corrections we have found are indeed negligible at small scales, except for the $\lambda Dp^{jk}/c^3$ term. It is necessary to retain this term because neglecting it leads to causality violation. Causality imposes a lower bound on $\lambda$ \cite{Muller99,BL99,KC24}. 

To see this, let us consider a plane front moving along the $x^3=z$ direction into a region where the fluid is at rest, so $\epsilon=v^j=p^{jk}=0$. The hydrodynamic variables are continuous at the front, while their $z-$ derivatives are not. If the front moves with speed $V$, any variable $X$ that is constant at the front obeys $X_{,t}+VX_{,z} =0$. Replacing the time derivatives into the equations of motion (assuming all variables depend on $z$ only) we get ($a,b=1,2$)

\be
0=-\frac1{\mu c^3}V\epsilon_{,z}+\frac{4}{3c}v^z_{,z}=0
\te

\be
0=-\frac4{c^2}Vv_{j,z}+\frac1{\mu\;c^2}\epsilon_{,j}+\frac3{c^2}p_{jz,z}
\te

\be
0=\frac1{8c}\left[v_{j,k}+v_{k,j}-\frac23v^3_{,3}\delta_{jk}\right]-\frac{\lambda}{c^3}Vp_{jk,z}
\te
We obtain a set of equations for the scalar quantities $\left(\epsilon,v^z,p^{zz}\right)$, another one for the vector quantities $\left(v^a,p^{az}\right)$ and a trivial equation for $p_{ab}=-p_{zz}\delta_{ab}/2$. The first set yields $V=0$ or else
a front speed $V_s$, where
\be
\left(\frac {V_s}c\right)^2=\frac13+\frac1{8\lambda}
\label{bound1}
\te
while  the second set yields a front speed $V_v$ given by

\be
\left(\frac {V_v}c\right)^2=\frac3{32\lambda}
\label{bound2}
\te
The strictest bound is given by the first set, $\lambda\ge 3/16$. Moreover, when this bound is approached, the speed of sound for the scalar variables approaches the speed of light, necessarily much higher than the fluid speed. We are justified in assuming the fluid is ``incompressible'', in the sense that $v^k_{,k}=p^{jk}_{,jk}=0$. $\epsilon$ becomes a Lagrange multiplier, enforcing the consistency of the momentum conservation equation.

We observe that in K41 theory the velocity of an eddy diminishes with eddy size. Therefore, a global subsonic flow will remain subsonic at all scales.  A fully relativistic flow, on the other hand, would necessarily be compressible. We discuss this point in section \ref{finalremarks}.

\section{Relativistic Green functions}\label{largec}

In this section, we discuss the retarded propagators and the correlation functions of the model. For the reasons discussed in the previous paragraph, we discard terms containing excess velocity fields, and take as our equations of motion

\be
0=Dv_j+\frac1{4\mu}\epsilon_{,j}+\frac34p_{jl,l}
\label{eq11}
\te
where it is assumed that $v^j_{,j}=0$, which determines the value of $\epsilon$, and

\be
0=\frac18\left[v_{j,k}+v_{k,j}-\frac23v^l_{,l}\delta_{jk}\right]+\frac3{32\eta}p_{jk}+\frac{\lambda}{c^2}Dp_{jk}
\label{eq21}
\te
where $\lambda\approx 1$. We also discard the scalar degrees of freedom in $p_{ij}$ and the non-propagating tensor mode \cite{HisLind3,GPEC,EC2,BD21}. This means that $p_{ij}=q_{i,j}+q_{j,i}$ with $q_{j,j}=0$. The projector on this tensor space is

\be
\Delta^{jk}_{lm}=\frac1{2k^2}\left[k^jk_l\Delta^k_m+k^jk_m\Delta^k_l+k^kk_l\Delta^j_m+k^kk_m\Delta^j_l\right]
\te
where

\be
\Delta^{jk}=\delta^{jk}-\frac{k^jk^k}{k^2}
\te
In the MSR approach, we must include two auxiliary fields, $a_j$ conjugated to $v^j$ and $b_{jk}$ conjugated to $p^{jk}$; $a_j$ has dimensions of $TL^{-4}$ and $b_{jk}$ of $L^{-3}$. We have $16$ two point functions, which split into $4$ physical field correlations, $4$ causal propagators, $4$ advanced propagators, and $4$ correlations among auxiliary fields, which are identically zero on-shell. We first investigate the retarded propagators. 

\subsection{Retarded propagators}\label{retprops}

Recall the Schwinger-Dyson-Wyld equations (\ref{props1}).
The equations for the retarded propagators are those in which both the indexes $\alpha$ and $\gamma$ denote auxiliary fields. We get

\bea
&&i\Delta^j_k\left(x,x'\right)\delta\left(t-t'\right)=\partial_t\left\langle v^j\left(x,t\right)a_k\left(x',t'\right)\right\rangle+\frac34\partial_l\left\langle p^{jl}\left(x,t\right)a_k\left(x',t'\right)\right\rangle\nn
&+&2\int d^3yds\left\{\frac{\delta \Gamma_Q}{\delta \left\langle a_j\left(x,t\right)v^l\left(y,s\right)\right\rangle}\left\langle v^l\left(y,s\right)a_k\left(x',t'\right)\right\rangle\right.\nn
&+&\left.\frac{\delta \Gamma_Q}{\delta \left\langle a_j\left(x,t\right)p^{lm}\left(y,s\right)\right\rangle}\left\langle p^{lm}\left(y,s\right)a_k\left(x',t'\right)\right\rangle\right\}
\label{eq1}
\tea

\bea
&&0=\left[\frac{\lambda}{c^2}\partial_t+\frac3{32\eta}\right]\left\langle p^{pq}\left(x,t\right)a_k\left(x',t'\right)\right\rangle+\frac18\left\langle \left[v_{p,q}+v_{q,p}\right]\left(x,t\right)a_k\left(x',t'\right)\right\rangle\nn
&+&2\int d^3yds\left\{\frac{\delta \Gamma_Q}{\delta \left\langle b_{pq}\left(x,t\right)v^l\left(y,s\right)\right\rangle}\left\langle v^l\left(y,s\right)a_k\left(x',t'\right)\right\rangle\right.\nn
&+&\left.\frac{\delta \Gamma_Q}{\delta \left\langle b_{pq}\left(x,t\right)p^{lm}\left(y,s\right)\right\rangle}\left\langle p^{lm}\left(y,s\right)a_k\left(x',t'\right)\right\rangle\right\}
\label{eq2}
\tea

\bea
&&0=\partial_t\left\langle v^j\left(x,t\right)b_{kn}\left(x',t'\right)\right\rangle+\frac34\partial_l\left\langle p^{jl}\left(x,t\right)b_{kn}\left(x',t'\right)\right\rangle\nn
&+&2\int d^3yds\left\{\frac{\delta \Gamma_Q}{\delta \left\langle a_j\left(x,t\right)v^l\left(y,s\right)\right\rangle}\left\langle v^l\left(y,s\right)b_{kn}\left(x',t'\right)\right\rangle\right.\nn
&+&\left.\frac{\delta \Gamma_Q}{\delta \left\langle a_j\left(x,t\right)p^{lm}\left(y,s\right)\right\rangle}\left\langle p^{lm}\left(y,s\right)b_{kn}\left(x',t'\right)\right\rangle\right\}
\label{eq3}
\tea

\bea
&&i\Delta^{pq}_{kn}\left(x,x'\right)\delta\left(t-t'\right)\nn
&=&\left[\frac{\lambda}{c^2}\partial_t+\frac3{32\eta}\right]\left\langle p^{pq}\left(x,t\right)b_{kn}\left(x',t'\right)\right\rangle+\frac18\left\langle \left[v_{p,q}+v_{q,p}\right]\left(x,t\right)b_{kn}\left(x',t'\right)\right\rangle\nn
&+&2\int d^3yds\left\{\frac{\delta \Gamma_Q}{\delta \left\langle b_{pq}\left(x,t\right)v^l\left(y,s\right)\right\rangle}\left\langle v^l\left(y,s\right)b_{kn}\left(x',t'\right)\right\rangle\right.\nn
&+&\left.\frac{\delta \Gamma_Q}{\delta \left\langle b_{pq}\left(x,t\right)p^{lm}\left(y,s\right)\right\rangle}\left\langle p^{lm}\left(y,s\right)b_{kn}\left(x',t'\right)\right\rangle\right\}
\label{eq4}
\tea
Let us start with eqs. (\ref{eq1}) and (\ref{eq2}).
Assuming spherical symmetry, we parameterize

\bea
&&\left\langle v^j\left(x,t\right)a_k\left(x',t'\right)\right\rangle=\int\frac{d^3k}{\left(2\pi\right)^3}\;e^{ik\left(x-x'\right)}\;\Delta^j_k\left[k\right]G_{VA}\left[k,t-t'\right]\nn
&&\left\langle p^{jl}\left(x,t\right)a_k\left(x',t'\right)\right\rangle=i\int\frac{d^3k}{\left(2\pi\right)^3}\;e^{ik\left(x-x'\right)}\;\Delta^{jl,k}\left[k\right]G_{PA}\left[k,t-t'\right]\nn
&&\left\langle v^j\left(x,t\right)b_{mn}\left(x',t'\right)\right\rangle=i\int\frac{d^3k}{\left(2\pi\right)^3}\;e^{ik\left(x-x'\right)}\;\Delta^{mn,j}\left[k\right]G_{VB}\left[k,t-t'\right]\nn
&&\left\langle p^{jl}\left(x,t\right)b_{mn}\left(x',t'\right)\right\rangle=\int\frac{d^3k}{\left(2\pi\right)^3}\;e^{ik\left(x-x'\right)}\;\Delta^{jl}_{mn}G_{PB}\left[k,t-t'\right]
\tea
where $G_{VA}$ is dimensionless, $G_{VB}$ and $G_{PA}$ have dimensions of $LT^{-1}$, $G_{PB}$ has dimensions of $L^2T^{-2}$ and

\be
\Delta^{jl,k}\left[k\right]=\frac1k\left[k^l\Delta^{jk}\left[k\right]+k^j\Delta^{lk}\left[k\right]\right]
\te
We also introduce the scalar self-energies

\bea
&&\frac{\delta \Gamma_Q}{\delta \left\langle a_j\left(x,t\right)v^l\left(y,s\right)\right\rangle}=\frac12\int\frac{d^3k}{\left(2\pi\right)^3}\;e^{ik\left(x-x'\right)}\;\Delta^j_l\left[k\right]\Sigma_{AV}\left[k,t-t'\right]\nn
&&\frac{\delta \Gamma_Q}{\delta \left\langle a_j\left(x,t\right)p^{lm}\left(y,s\right)\right\rangle}=\frac i2\int\frac{d^3k}{\left(2\pi\right)^3}\;e^{ik\left(x-x'\right)}\;\Delta^{lm,j}\left[k\right]\Sigma_{AP}\left[k,t-t'\right]\nn
&&\frac{\delta \Gamma_Q}{\delta \left\langle b_{pq}\left(x,t\right)v^l\left(y,s\right)\right\rangle}=\frac i2\int\frac{d^3k}{\left(2\pi\right)^3}\;e^{ik\left(x-x'\right)}\;\Delta^{pq,l}\left[k\right]\Sigma_{BV}\left[k,t-t'\right]\nn
&&\frac{\delta \Gamma_Q}{\delta \left\langle b_{pq}\left(x,t\right)p^{lm}\left(y,s\right)\right\rangle}=\frac12\int\frac{d^3k}{\left(2\pi\right)^3}\;e^{ik\left(x-x'\right)}\;\Delta^{pq,lm}\left[k\right]\Sigma_{BP}\left[k,t-t'\right]
\tea
$\Sigma_{AV}$ has units of $T^{-2}$, $\Sigma_{AP}$ and $\Sigma_{BV}$ of $L^{-1}T^{-1}$, and $\Sigma_{BP}$ of $L^{-2}$. Then eq. (\ref{eq1}) becomes 

\bea
&&i\delta\left(t-t'\right)=\partial_tG_{VA}\left[k,t-t'\right]-\frac34kG_{PA}\left[k,t-t'\right]\nn
&+&\int ds\left\{\Sigma_{AV}\left[k,t-s\right]G_{VA}\left[k,s-t'\right]-2\Sigma_{AP}\left[k,t-s\right]G_{PA}\left[k,s-t'\right]\right\}
\tea
and eq. (\ref{eq2})

\bea
&&0=\left[\partial_t+\frac{3c^2}{32\lambda\eta}\right]G_{PA}\left[k,t-t'\right]+\frac{c^2k}{8\lambda}G_{VA}\left[k,t-t'\right]\nn
&+&\frac{c^2}{\lambda}\int ds\left\{\Sigma_{BV}\left[k,t-s\right]G_{VA}\left[k,s-t'\right]+\Sigma_{BP}\left[k,t-s\right]G_{PA}\left[k,s-t'\right]\right\}
\tea
We make the Markovian approximation (for a recent discussion of this approximation, see \cite{Canet22})

\be
\Sigma_{XY}\left[k,t-s\right]\approx\delta\left(t-s\right)\bar\Sigma_{XY}\left[k\right]
\label{Markovianapproximation}
\te
where

\be
\bar\Sigma_{XY}\left[k\right]=\int_0^{\infty}dt\;\Sigma_{XY}\left[k,t\right]
\te
We finally get the system

\be 
\partial_t\left(\begin{array}{c}G_{VA}\\G_{PA}\end{array}\right)+{\mathbf\Sigma}\left(\begin{array}{c}G_{VA}\\G_{PA}\end{array}\right)=\left(\begin{array}{c}i\delta\left(t\right)\\0\end{array}\right)
\te
where

\be
{\mathbf\Sigma}=\left(\begin{array}{cc}\bar\Sigma_{AV}&-\frac34k\left(1+\frac8{3k}\bar\Sigma_{AP}\right)\\\frac{c^2k}{8\lambda}\left[1+\frac8k\bar\Sigma_{BV}\right]&\frac{3c^2}{32\lambda\eta}\left[1+\frac{32\eta}3\bar\Sigma_{BP}\right]\end{array}\right)
\te
We have two eigenvalues $\sigma_{\pm}$ and the corresponding right eigenvectors

 \be
\mathbf{S}_{\pm}=\left(\begin{array}{c} s_{\pm}\\-1\end{array}\right)
\te
If the self-energies do not diverge as $c^2\to\infty$, then one of the eigenvalues, say $\sigma_+$, must diverge as $c^2$, but then the corresponding $s_+\propto c^{-2}$. To leading order

\bea
&&\sigma_+=\frac{3c^2}{32\lambda\eta}\left[1+\frac{32}3\eta\bar\Sigma_{BP}\right]\nn
&&s_+=8\frac{\lambda\eta k}{c^2}\frac{\left(1+\frac 4{3k}\bar\Sigma_{AP}\right)}{\left[1+\frac{32}3\eta\bar\Sigma_{BP}\right]}
\label{smas}
\tea
The $\sigma_-$ eigenvalue remains finite as $c\to\infty$, so to LO it must be

\bea
&&s_-=\frac{3}{4\eta k}\frac{\left[1+\frac{32}3\eta\bar\Sigma_{BP}\right]}{\left[1+\frac8k\bar\Sigma_{BV}\right]}\nn
&&\sigma_-=\bar\Sigma_{AV}+\eta k^2\frac{\left(1+\frac 4{3k}\bar\Sigma_{AP}\right)\left[1+\frac8k\bar\Sigma_{BV}\right]}{\left[1+\frac{32}3\eta\bar\Sigma_{BP}\right]}
\label{sigmamenos}
\tea
$s_{\pm}$ have units of $TL^{-1}$. To finish the job, we decompose

\be
\left(\begin{array}{c}1\\0\end{array}\right)=\frac1{s_--s_+}\left[-\left(\begin{array}{c}s_+\\-1\end{array}\right)+\left(\begin{array}{c}\bar s_-\\-1\end{array}\right)\right]
\te
leading to

\bea
G_{VA}&=&i\left[g_{VA}^+\;e^{-\sigma_{+}\left(t-t'\right)}+g_{VA}^-\;e^{-\sigma_{-}\left(t-t'\right)}\right]\theta\left(t-t'\right)\nn
G_{PA}&=&i\left[g_{PA}^+\;e^{-\sigma_{+}\left(t-t'\right)}+g_{PA}^-\;e^{-\sigma_{-}\left(t-t'\right)}\right]\theta\left(t-t'\right)
\tea
where

\bea
g_{VA}^+&=&1-g_{VA}^-=\frac{-s_+}{s_--s_+}\nn
g_{PA}^+&=&\frac{1}{s_--s_+};\;\;g_{VA}^-=\frac{-1}{s_--s_+} \nn
\tea
We now consider the second pair of equations, eqs. (\ref{eq3}) and (\ref{eq4}), which become

\be
0=\partial_tG_{VB}+\bar\Sigma_{AV}G_{VB}+\frac38k\left[1+\frac{4}{3k}\bar\Sigma_{AP}\right]G_{PB}
\te

\be
i\delta\left(t-t'\right)=\left[\frac{\lambda}{c^2}\partial_t+\frac3{32\eta}\right]G_{PB}-\frac14kG_{VB}-2\bar\Sigma_{BV}G_{VB}+\bar\Sigma_{BP}G_{PB}
\te
In matrix form

\be 
\partial_t\left(\begin{array}{c}G_{VB}\\G_{PB}\end{array}\right)+{\mathbf\Sigma}'\left(\begin{array}{c}G_{VB}\\G_{PB}\end{array}\right)=\frac{ic^2}{\lambda}\left(\begin{array}{c}0\\1\end{array}\right)\delta\left(t\right)
\label{second}
\te
where

\be
{\mathbf\Sigma}'=\left(\begin{array}{cc}\bar\Sigma_{AV}&\frac38k\left[1+\frac{4}{3k}\bar\Sigma_{AP}\right]\\-\frac14\frac{c^2k}{\lambda}\left[1+\frac8k\bar\Sigma_{BV}\right]&\frac{3c^2}{32\lambda\eta}\left[1+\frac{32}3\eta\bar\Sigma_{BP}\right]\end{array}\right)
\te
The eigenvalues are once again $\sigma_{\pm}$, and the right eigenvectors

 \be
\mathbf{S}'_{\pm}=\left(\begin{array}{c} \frac12s_{\pm}\\ 1\end{array}\right)
\te
We now write

\be
\left(\begin{array}{c}0\\1\end{array}\right)=\frac1{s_--s_+}\left[-s_+\left(\begin{array}{c}\frac12s_-\\1\end{array}\right)+s_-\left(\begin{array}{c}\frac12s_+\\1\end{array}\right)\right]
\te
and therefore

\bea
G_{VB}&=&i\left[g_{VB}^+\;e^{-\sigma_{+}\left(t-t'\right)}+g_{VB}^-\;e^{-\sigma_{-}\left(t-t'\right)}\right]\theta\left(t-t'\right)\nn
G_{PB}&=&i\left[g_{PB}^+\;e^{-\sigma_{+}\left(t-t'\right)}+g_{PB}^-\;e^{-\sigma_{-}\left(t-t'\right)}\right]\theta\left(t-t'\right)
\tea
where

\bea
g_{VB}^+&=&-g_{VB}^-=\frac{c^2}{2\lambda}\frac{s_-s_+}{s_--s_+}\nn
g_{PB}^+&=&\frac{c^2}{\lambda}-g_{PB}^-=\frac{c^2}{\lambda}\frac{s_-}{s_--s_+}
\tea

\subsection{Correlation functions}\label{thecorrelators}

We may now compute the correlations. They are given by eq. (\ref{Hadamard}), where the variation of the 2PIEA is written in terms of the noise kernel as in eq. (\ref{noise}). As usual, we expand

\bea
&&\left\langle v^j\left(x,t\right)v^k\left(x',t'\right)\right\rangle=\int\frac{d^3k}{\left(2\pi\right)^3}\;e^{ik\left(x-x'\right)}\;\Delta^{jk}\left[k\right]G_{VV}\left[k,t-t'\right]\nn
&&\left\langle p^{jl}\left(x,t\right)v^k\left(x',t'\right)\right\rangle=i\int\frac{d^3k}{\left(2\pi\right)^3}\;e^{ik\left(x-x'\right)}\;\Delta^{jl,k}G_{PV}\left[k,t-t'\right]\nn
&&\left\langle p^{jl}\left(x,t\right)p^{km}\left(x',t'\right)\right\rangle=\int\frac{d^3k}{\left(2\pi\right)^3}\;e^{ik\left(x-x'\right)}\;\Delta^{jl,km}G_{PV}\left[k,t-t'\right]
\tea
We also write the Fourier transforms of the noise kernels as

\bea
N^{lm}_{AA}\left[k,t-t'\right]&=&\Delta^{lm}\bar N_{AA}\left[k\right]\delta\left(t-t'\right)\nn
N^{l,mn}_{AB}\left[k,t-t'\right]&=&-N_{BA}^{mn,l}\left[k,t-t'\right]=i\Delta^{mn,l}\bar N_{AB}\left[k\right]\delta\left(t-t'\right)\nn
N^{ln,mp}_{BB}\left[k,t-t'\right]&=&\Delta^{ln,mp}\bar N_{BB}\left[k\right]\delta\left(t-t'\right)
\label{noisedefs}
\tea
A stochastic force on the right hand side of eq. (\ref{eq11}) would have units of acceleration $LT^{-2}$ and one on the right hand side of eq. (\ref{eq21}) of inverse time $T^{-1}$. It follows that $\bar N^{jk}_{AA}\left[k\right]$ has units of $L^5T^{-3}$, consistent with the K41 estimate $\bar N_{AA}\propto\epsilon k^{-3}$. $\bar N_{BA}\left[k\right]$ has units of $L^4T^{-2}$ and $\bar N_{BB}\left[k\right]$ of $L^3T^{-1} $. 

Let us start with

\bea
\left\langle v^j\left(x,t\right)v^k\left(x',t'\right)\right\rangle&=&-\int d^3ydsd^3y'ds'\nn
&&\left\{G^j_{VAl}\left(x,t;y,s\right)G^k_{VAm}\left(x',t';y',s'\right)N^{lm}_{AA}\left(y,s;y',s'\right)\right.\nn
&+&G^j_{VAl}\left(x,t;y,s\right)G^k_{VBmn}\left(x',t';y',s'\right)N^{l,mn}_{AB}\left(y,s;y',s'\right)\nn
&+&G^j_{VBln}\left(x,t;y,s\right)G^k_{VAm}\left(x',t';y',s'\right)N^{ln,m}_{BA}\left(y,s;y',s'\right)\nn
&+&\left.G^j_{VBln}\left(x,t;y,s\right)G^k_{VBmp}\left(x',t';y',s'\right)N^{ln,mp}_{BB}\left(y,s;y',s'\right)\right\}
\tea
Or else, Fourier transforming (note the third term)

\bea
&&G_{VV}^{jk}\left[k,t-t'\right]=-\int dsds'\;\left\{G^j_{VAl}\left(k,t-s\right)G^k_{VAm}\left(-k,t'-s'\right)N^{lm}_{AA}\left(k,s-s'\right)\right.\nn
&+&G^j_{VAl}\left(k,t-s\right)G^k_{VBmn}\left(-k,t'-s'\right)N^{l,mn}_{AB}\left(k,s-s'\right)\nn
&+&G^j_{VBln}\left(k,t-s\right)G^k_{VAm}\left(-k,t'-s'\right)N^{m,ln}_{AB}\left(-k,s-s'\right)\nn
&+&\left.G^j_{VBln}\left(k,t-s\right)G^k_{VBmp}\left(-k,t'-s'\right)N^{ln,mp}_{BB}\left(k,s-s'\right)\right\}
\tea
Separate the tensor part and perform the algebra

\bea
&&G_{VV}\left[k,t-t'\right]=-\int dsds'\;\left\{G_{VA}\left(k,t-s\right)G_{VA}\left(k,t'-s'\right)N_{AA}\left(k,s-s'\right)\right.\nn
&+&2G_{VA}\left(k,t-s\right)G_{VB}\left(k,t'-s'\right)N_{AB}\left(k,s-s'\right)\nn
&+&2G_{VB}\left(k,t-s\right)G_{VA}\left(k,t'-s'\right)N_{AB}\left(k,s-s'\right)\nn
&+&\left.2G_{VB}\left(k,t-s\right)G_{VB}\left(k,t'-s'\right)N_{BB}\left(k,s-s'\right)\right\}
\tea
When $t\ge t'$

\bea
&&G_{VV}\left[k,t-t'\right]=e^{-\sigma_+\left(t-t'\right)}\left\{\left(g^+_{VA}\bar N_{AA}+2g^+_{VB}\bar N_{AB}\right)\left[\frac{g_{VA}^+}{2\sigma_+}+\frac{g_{VA}^-}{\sigma_++\sigma_-}\right]\right.\nn
&+&\left.2\left(g^+_{VA}\bar N_{AB}+g^+_{VB}\bar N_{BB}\right)\left[\frac{g_{VB}^+}{2\sigma_+}+\frac{g_{VB}^-}{\sigma_++\sigma_-}\right]\right\}\nn
&+&e^{-\sigma_-\left(t-t'\right)}\left\{\left(g^-_{VA}\bar N_{AA}+2g^-_{VB}\bar N_{AB}\right)\left[\frac{g_{VA}^+}{\sigma_++\sigma_-}+\frac{g_{VA}^-}{2\sigma_-}\right]\right.\nn
&+&\left.2\left(g^-_{VA}\bar N_{AB}+g^-_{VB}\bar N_{BB}\right)\left[\frac{g_{VB}^+}{\sigma_++\sigma_-}+\frac{g_{VB}^-}{2\sigma_-}\right]\right\}
\tea
Let us define

\bea
&&\frac{g^+_{VA}}{g^-_{VB}}=\frac{2\lambda}{c^2s_-}\equiv \gamma\approx O\left(c^{-2}\right)\nn
&&\frac{1-g^+_{VA}}{g^-_{VB}}=-\frac{2\lambda}{c^2s_+}\equiv \gamma'\approx O\left(1\right)
\tea
Then

\bea
&&G_{VV}\left[k,t-t'\right]=2\left(g^{-}_{VB}\right)^2e^{-\sigma_+\left(t-t'\right)}\left\{\left(\frac12\gamma\bar N_{AA}-\bar N_{AB}\right)\left[\frac{\gamma}{2\sigma_+}+\frac{\gamma'}{\sigma_++\sigma_-}\right]\right.\nn
&-&\left.\left(\gamma\bar N_{AB}-\bar N_{BB}\right)\left[\frac{1}{2\sigma_+}-\frac{1}{\sigma_++\sigma_-}\right]\right\}\nn
&+&2\left(g^{-}_{VB}\right)^2e^{-\sigma_-\left(t-t'\right)}\left\{\left(\frac12\gamma'\bar N_{AA}+\bar N_{AB}\right)\left[\frac{\gamma}{\sigma_++\sigma_-}+\frac{\gamma'}{2\sigma_-}\right]\right.\nn
&-&\left.\left(\gamma'\bar N_{AB}+\bar N_{BB}\right)\left[\frac{1}{\sigma_++\sigma_-}-\frac{1}{2\sigma_-}\right]\right\}
\tea
We perform a similar analysis for the other correlations

\bea
&&G_{VP}\left[k,t-t'\right]=g^-_{VB}e^{-\sigma_+\left(t-t'\right)}\left\{-2g_{PA}^+\left(\frac12\gamma\bar N_{AA}-\bar N_{AB}\right)\left[\frac{1}{2\sigma_+}-\frac{1}{\sigma_++\sigma_-}\right]\right.\nn
&+&\left.\left(\gamma\bar N_{AB}-\bar N_{BB}\right)\left[\frac{g_{PB}^+}{2\sigma_+}+\frac{g_{PB}^-}{\sigma_++\sigma_-}\right]\right\}\nn
&+&g^-_{VB}e^{-\sigma_-\left(t-t'\right)}\left\{-2g_{PA}^+\left(\frac12\gamma'\bar N_{AA}+\bar N_{AB}\right)\left[\frac{1}{\sigma_++\sigma_-}-\frac{1}{2\sigma_-}\right]\right.\nn
&+&\left.\left(\gamma'\bar N_{AB}+\bar N_{BB}\right)\left[\frac{g_{PB}^+}{\sigma_++\sigma_-}+\frac{g_{PB}^-}{2\sigma_-}\right]\right\}
\label{gvp}
\tea
Observe that $g_{PA}^+=\gamma g^+_{PB}/2$. 

\bea
&&G_{PP}\left[k,t-t'\right]=\frac{c^2}{\lambda}\gamma'g_{VB}^-e^{-\sigma_+\left(t-t'\right)}\left\{\gamma g_{PB}^+\left(\frac12\gamma\bar N_{AA}-\bar N_{AB}\right)\left[\frac{1}{2\sigma_+}-\frac{1}{\sigma_++\sigma_-}\right]\right.\nn
&+&\left.\left(-\gamma\bar N_{AB}+\bar N_{BB}\right)\left[\frac{g_{PB}^+}{2\sigma_+}+\frac{g_{PB}^-}{\sigma_++\sigma_-}\right]\right\}\nn
&+&\frac{c^2}{\lambda}\gamma g_{VB}^-e^{-\sigma_-\left(t-t'\right)}\left\{-\frac12\gamma g_{PB}^+\left(\gamma'\bar N_{AA}+2\bar N_{AB}\right)\left[\frac{1}{\sigma_++\sigma_-}-\frac{1}{2\sigma_-}\right]\right.\nn
&+&\left.\left(\gamma'\bar N_{AB}+\bar N_{BB}\right)\left[\frac{g_{PB}^+}{\sigma_++\sigma_-}+\frac{g_{PB}^-}{2\sigma_-}\right]\right\}
\label{gpp}
\tea

\subsection{Recovering K41}\label{RTP}

We see from the previous section that the flow of a relativistic fluid is in general the superposition of slowly decaying modes, decaying approximately at the non relativistic rate, and fast decaying modes, whose decay rate scales as $c^2$. There are therefore two basic flow patterns, one in which there are only slowly decaying modes and the other in which there are only fast decaying modes. We ask now whether the former merges smoothly with the Kolmogorov cascade as $c^2\to\infty$, as claimed in \cite{FOz10}.

We see from eqs. (\ref{gvp}) and (\ref{gpp}) that if we impose the relations

\bea
&&\bar N_{AB}=\frac12\gamma\bar N_{AA}\nn
&&\bar N_{BB}=\frac12\gamma^2\bar N_{AA}
\label{low2}
\tea
The fast decaying modes drop out entirely, yielding the correlation functions

\bea
G_{VV}\left[k,t-t'\right]&=&\frac{\bar N_{AA}}{2\sigma_-}e^{-\sigma_-\left|t-t'\right|}\nn
G_{VP}\left[k,t-t'\right]&=&\frac1{s_-}\frac{\bar N_{AA}}{2\sigma_-}e^{-\sigma_-\left|t-t'\right|}\nn
G_{PP}\left[k,t-t'\right]&=&\frac2{s_-^2}\frac{\bar N_{AA}}{2\sigma_-}\;e^{-\sigma_-\left|t-t'\right|}
\label{slowcorrelations}
\tea
Remarkably, this is consistent with the observation that, in the limit where $c^2\to\infty$, we may regard the auxiliary field $b_{jk}$ as a Lagrange multiplier enforcing the constraint

\be
p_{jk}=-\frac43\eta\left(v_{j,k}+v_{k,j}\right)
\label{lowestorder}
\te 
This allows us to compute all correlations in terms of $G^{jk}_{VV}$. Factoring out the tensor parts, we get

\bea
G_{VP}&=&\frac43\eta kG_{VV}\nn
G_{PP}&=&2\left(\frac43\eta k\right)^2G_{VV}
\tea
To get eqs. (\ref{low2}) we replace $\left(4/3\right)\eta k$ by its ``dressed'' value $1/s_-$ (eq. (\ref{sigmamenos})), and compute the  noise kernels from the lowest order term in the 2PIEA

\bea
\bar N^{jk}_{AA}\left(x,x'\right)&=&\int\;dt\int\;d^yd^3y'\;\Delta^{jj'}\left(x-y\right)\Delta^{kk'}\left(x'-y'\right)\nn
&&\left\langle \left(v^mv^{j'}_{,m}\right)\left(y,t\right)\left(v^nv^{k'}_{,n}\right)\left(y',t'\right)\right\rangle_{2PI}
\label{selfconsis}
\tea 
and similarly for $N^{j,kl}_{AB}$ and $N^{jk,lm}_{BB}$.

If we assume a scaling solution where $\bar N_{AA}\propto k^{-\beta}$ and $\sigma_-\propto k^{\alpha}$, then from eq. (\ref{selfconsis}) we get 

\be
3\alpha=5-\beta
\label{Kscaling}
\te 
K41 becomes a particular solution, associated to exponents $\beta=3$ and $\alpha =2/3$, and the relationships

\bea
&&\bar N_{AA}=2\mathcal{D}C^{3/2}_K\epsilon\frac1{k^3}\nn
&&\sigma_-\approx\bar\Sigma_{VA}=\mathcal{D}C_K^{1/2}\left(\epsilon k^2\right)^{1/3}
\label{scalingsigmamenos}
\tea
where $C_K$ the (possibly dressed) Kolmogorov constant \cite{Sre95}, and $\mathcal{D}$ is a numerical factor.

The existence of this cascade has been pointed out in 
\cite{FOz10}, but in a model that is unstable and violates causality. We have shown here that the same cascade obtains in causal and  stable models.

There is a point which must be clarified, however. The requirement that we recover K41 in the nonrelativistic limit leads us to identify the decay rate of the slow modes with the Kolmogorov value $\sigma_-\propto (\epsilon k^2)^{1/3}$, but by doing so we are ignoring the fact that the sweeping of small eddies by large ones yields instead a larger decay rate $\propto Vk$, where $V$ is the large scale velocity. This effect has been discussed  by Kraichnan \cite{Kr64} and verified in numerical and renormalization group analysis \cite{CK89,SS92,DJLW17,TCW18,Tar20,Canet22,Gor21}. 

In the non relativistic literature, the correct decay rate is found by first formulating the theory within a Lagrangian or quasi-Lagrangian frame
\cite{LB87,KR65,Kr66,Gaw97,LvL93}. In this frame the Kolmogorov decay rate obtains, but it is converted into the advection decay rate when translated into an Eulerian frame. Nevertheless, the equal time correlations remain at their Kolmogorov value. We are not aware of a similar treatment in the relativistic case, we shall return to this point in the final remarks.

\subsection{The fast decaying flow pattern}

If instead of eq. (\ref{low2}) we request

\bea
\bar N_{AB}&=&-\frac12\gamma'\bar N_{AA}\nn
\bar N_{BB}&=&-\gamma'\bar N_{AB}
\label{fastnoises}
\tea
then the slowly decaying contibutions to the correlators cancel out and we obtain 

\bea
G^{\left(FD\right)}_{VV}\left[k,t-t'\right]&=&\frac{\bar N_{AA}}{2\sigma_+}e^{-\sigma_+\left|t-t'\right|}\nn
G^{\left(FD\right)}_{VP}\left[k,t-t'\right]&=&\frac1{s_+}\frac{\bar N_{AA}}{2\sigma_+}e^{-\sigma_+\left|t-t'\right|}\nn
G^{\left(FD\right)}_{PP}\left[k,t-t'\right]&=&\frac2{s_+^2}\frac{\bar N_{AA}}{2\sigma_+}\;e^{-\sigma_+\left|t-t'\right|}
\label{fastcorrelations}
\tea
Thus obtaining a flow pattern made up of fast decaying modes only.

It is interesting to observe that sweeping effects are nor relevant regarding the fast decaying modes. Indeed for modes above the Kolmogorov scale $k\ge\delta^{-1}\approx (\epsilon\eta^{-3})^{1/4}$ we find

\be
\frac{Vk}{\sigma_+}\approx \left(\frac{V^2}{c^2}\right)\mathbf{Re}^{-1/4}\ll 1
\te
We may ask further if these relations admit scaling solutions analogous to the Kolmogorov one. Recall that $\sigma_+$ is independent of $k$ to leading order and assume $s_+\propto k^{\delta}$, so that $\gamma'\propto k^{-\delta}$. Then these equations are compatible with a scaling solution where $\bar N_{AA}\propto  k^{-\beta}$ and $G^{\left(FD\right)}_{VV}\propto k^{-\beta}$, provided that

\be
 k^{-\beta}\propto k^{5-2\beta}
\te
We conclude that $\beta=5$, and then $G_{PP}\propto  k^{-2\delta-5}$. In the same way that in the slowly decaying case we found eq. (\ref{Kscaling}) but could not identify the $\beta$ and $\alpha$ exponents without further assumptions, here we found the $\beta$ exponent but need further assumptions to identify $\delta$. As a motivation for further work, we shall discuss possible such assumptions in the Appendix.

\section{Final remarks}\label{finalremarks}

In this work, we have attempted to take advantage of our deeper understanding of non-relativistic turbulence to investigate its relativistic counterpart. 

As a first step towards this goal, we have discussed a model of a massless relativistic fluid. We assume that the energy density diverges as $c^2$ as $c\to\infty$, but not faster. Under this hypothesis, the non relativistic limit of this system is an incompressible fluid obeying the NSE \cite{FOz08}. We have developed a simplified model that retains the properties of causality, Lorentz invariance and thermodynamic stability. This is possibly the main contribution of this work.

Our analysis reveals a certain duality in relativistic flows. Under the Markovian approximation, each mode is associated to two different decay rates. One of them converges to the nonrelativistic value, while the other diverges when $c\to\infty$. We therefore identify the former as slowly decaying, and the latter as fast decaying. We show that in general the correlators of the theory contain modes of both kinds. However, if we assume the relations eq. (\ref{low2}), the fast modes cancel out and we get a flow pattern with a definite nonrelativistic limit. It is natural to expect that this limit corresponds to K41 theory, and we point out the supplementary conditions under which this obtains. The existence of a Kolmogorov cascade in Landau-Lifshitz hydrodynamics is known \cite{FOz10}; we have generalized this result to a causal, stable model of relativistic hydrodynamics.

Similarly, if we assume the relations eq. (\ref{fastnoises}), then we obtain a flow built from fast decaying modes only. These relations admit a variety of scaling solutions, and this opens the issue of whether there is a set of supplementary conditions which would single a preferred one. As a motivation for future work, we present an example in the Appendix.

We stress that we are still very far from a full understanding of relativistic turbulence. Among several avenues that should be pursued we mention relaxing the incompressibility assumption, and allowing for sweeping effects in the two time correlations. The fulfillment  of these challenges will most likely involve non perturbative methods \cite{LP96,BLP98,LP98,Friedrich}, such as the renormalization group 
\cite{yaorz86,Wett93,Eyink94,Tom97,Smith98,Giles,V98,Zhou10,Rentrop,Rischke22,AK24,Verma,ZC02,Rosenhaus,Canet22}. 

While we intend to address these issues in future work, let us state why we believe the results presented here will stand in more complete models. In the case of compressible turbulence (\cite{MY71,MY75}) there is a substantial body of work which shows the possibility of applying Kolmogorov inertial-range theory, see 
\cite{ED18b,PWP88,PPW94,Alu11,Alu13,BG14,STSV26}, and particularly \cite{SC18}. Ref. \cite{McKO07}  explicitly discusses the approximate incompressibility which obtains at small scales in an otherwise compressible flow.

Concerning the issue of sweeping effects, we have already noted that they are likely to be minor for the fast decaying modes, and for the slowly decaying ones, the evidence for Kolmogorov scaling of the equal time correlators is so overwhelming that there is little doubt that the final result is correct. Developing a fully consistent derivation is desirable but challenging. In the nonrelativistic case, a quasi-Lagrangian condition on the flow may be imposed by a sequence of Galilean transformations that leave the NSE invariant \cite{LB87}. In the relativistic case, the closest thing to a Lagrangian formulation would be using synchronous curvilinear coordinates for the fluid parcels (see \cite{LL4} and also \cite{CQ72,AC21}), leading to a totally different mathematical set up. To the best of our knowledge such problem is largely unexplored, at least within a Martin-Siggia-Rose framework.

However difficult, only when these tasks are accomplished the cogency of the non equilibrium field theory approach to turbulence shall be established beyond doubt.

\vspace{6pt}

\section*{Acknowledgments}

E. C. acknowledges financial support from Universidad de Buenos Aires through Grant No. UBACYT
20020170100129BA and CONICET Grant No. PIP2017/19:11220170100817CO.

\appendix

\section{Scale invariant fast decaying flow: a proposal}

As we have seen in the text, the conditions that exclude fast decaying modes from the correlators of the theory admit a plurality of scaling solutions. Supplementary conditions are needed to single out one of them. In this case, the Kolmogorov hypothesis of a scale invariant energy flow leads to K41 scaling 
(eq. (\ref{scalingsigmamenos})).

Similarly, the conditions which exclude the slowly decaying modes admit scaling solutions, but in this case it is not clear what the required supplementary conditions may be.

As a motivation for future work, we shall consider what kind of solution follows if we assume the existence of a scale invariant {entropy} flow. This choice is suggested by the work by Eyink and Drivas \cite{Eyink18b}, who raised the possibility that a relativistic real fluid could support an {entropy} cascade, besides the usual energy one. The goal of the analysis is to identify the dimensionful parameter (called $\kappa$ below) that plays the role of $\epsilon$ in K41.

Let us begin by writing the entropy density eq. (\ref{sdens}) as

\be
s\approx\frac43\sigma_{SB}T^3\left[1-\frac32\frac{\lambda}{c^4}\left\langle p^{jk}p_{jk}\right\rangle\right]
\label{sdens2}
\te 
We are regarding $2\sigma_{SB}T^3$ as a constant and shall not make it explicit any longer.

From this point on, we follow the analysis of transport given by Kraichnan, see \cite{Kraichnan67}. In order to define an entropy spectrum, we consider the point-split entropy density 

\be
s\left(x,t\right)\propto\frac{\lambda}{c^4}\left\langle p^{jk}p_{jk}\right\rangle\left(x,t\right)\to \frac{\lambda}{c^4}s\left(x,x',t\right)\propto\left\langle p^{jk}\left(x,t\right)p_{jk}\left(x',t\right)\right\rangle
\te
The transport part of the entropy equation reads

\be
\left.\frac{\partial s}{\partial t}\right|_{\mathrm{transport}}=2\frac{\lambda}{c^4}\left\langle p^{jk}\left(x,t\right)\left(v^lp_{jk,l}\right)\left(x',t\right)\right\rangle
\te
Introducing the Fourier transforms $s_p$, $v^j_p$ and $p^{jk}_p$ we get

\be
\left.\frac{\partial s_p}{\partial t}\right|_{\mathrm{transport}}=\frac{\lambda}{c^4}\int \frac{d^3q}{\left(2\pi\right)^3}\frac{d^3r}{\left(2\pi\right)^3}h\left[\vec p,\vec q,\vec r\right]
\te 
where

\be
h\left[\vec p,\vec q,\vec r\right]=-ip_l\left[\left\langle p^{jk}_pv^l_qp^{jk}_r\right\rangle+\left\langle p^{jk}_pv^l_rp^{jk}_q\right\rangle\right]
\te
Observe that $h\left[\vec p,\vec q,\vec r\right]\propto\delta\left(\vec p+\vec q+\vec r\right)$. Moreover,

\be
h\left[\vec p,\vec q,\vec r\right]+h\left[\vec q,\vec r,\vec p\right]+h\left[\vec r,\vec p,\vec q\right]=0
\te
which implies that

\be
\int \frac{d^3p}{\left(2\pi\right)^3}\left.\frac{\partial s_p}{\partial t}\right|_{\mathrm{transport}}=0
\te
The entropy transported to modes above $p_0$ is then $\left(\lambda/c^4\right)H\left[p_0\right]$, where

\be
H\left[p_0\right]=H_+\left[p_0\right]-H_-\left[p_0\right]
\te
where

\bea
&&H_+\left[p_0\right]=\int_{p>p_0,q<p_0,r<p_0} \frac{d^3p}{\left(2\pi\right)^3}\frac{d^3q}{\left(2\pi\right)^3}\frac{d^3r}{\left(2\pi\right)^3}h\left[\vec p,\vec q,\vec r\right]\nn
&&H_-\left[p_0\right]=\int_{p<p_0,q>p_0,r>p_0} \frac{d^3p}{\left(2\pi\right)^3}\frac{d^3q}{\left(2\pi\right)^3}\frac{d^3r}{\left(2\pi\right)^3}h\left[\vec p,\vec q,\vec r\right]
\tea
To find the scaling of $H_+$ and $H_-$ we use the leading order result \cite{FGB94,FOK,FKO24,Zhou}

\bea
&&\left\langle p^{jk}\left[x,t\right]v^l\left[x',t\right]p^{jk}\left[x'',t\right]\right\rangle=\frac{-i\lambda}{c^2}\int\;d^3yds\nn
&&\left\langle p^{jk}\left[x,t\right]v^l\left[x',t\right]p^{jk}\left[x'',t\right]\left(b_{mn,p}v^pp^{mn}\right)\left(y,s\right)\right\rangle_{2PI}
\tea
Fourier transforming we get

\bea
&&\left\langle p_p^{jk}v_q^lp^{jk}_r\right\rangle\left(t\right)=\frac{i\lambda}{c^2}\delta\left(\vec p+\vec q+\vec r\right)\int\;ds\nn
&&\left\{p_lG_{PB}^{jk,mn}\left[p,t-s\right]G_{VV}^{lp}\left[q,t-s\right]G_{PP}^{jk,mn}\left[r,t-s\right]+\ldots\right\}
\tea
where $\ldots$ stands for all the other permutations of the fields. Since we only want the scaling of the entropy flux, it is enough to consider only this term, whereby $\left\langle p_p^{jk}\left(t\right)v_q^l\left(t\right)p^{jk}_r\left(t\right)\right\rangle\propto p^{-12-2\delta}$, and

\be
H_+\propto p^{-2-2\delta}
\te
This means that to get a scale invariant entropy flux we need 

\be
s_+\propto p^{-1}
\label{scalingsmas}
\te 
Observe that then $\bar N_{BB}$ scales as $p^{-3}$. 

Let us call $\kappa$ the scale invariant value of $H_+$. $\kappa$ has units of $L^4T^{-5}$. To recover the $p^{-5}$ scaling, we must have 

\be
G_{VV}\propto \frac{\kappa^2}{\left(c^2\right)^4}\frac1{p^5}
\te
Similarly, since $s_+$ has units of $TL^{-1}$, to recover the $p^{-1}$ scaling it must be $s_+\propto \kappa\left(\left(c^2\right)^3p\right)^{-1}$, and then

\be
G_{PP}\propto \left(c^2\right)^2\frac1{p^3}
\te 
so the entropy spectrum comes out scale invariant, and independent of $c$.

Our analysis shows that the hypothesis of a scale invariant entropy flow may single out a self similar solution out of the family consistent with eq. (\ref{fastcorrelations}), but it is not a proof of the actual existence of such solutions, much less of their phenomenological relevance. For example, we have not shown that the relevant self energies and noise kernels can actually be computed from suitable Feynman graphs. It is presented here only as a motivation for further inquiry.

%\conflictsofinterest{The authors declare no conflicts of interest. The funders had no role in the design of the study; in the collection, analyses, or interpretation of data; in the writing of the manuscript; or in the decision to publish the results.} 

%%%%%%%%%%%%%%%%%%%%%%%%%%%%%%%%%%%%%%%%%%

%\reftitle{References}

\end{document}